# EXPERIMENTAL INVESTIGATION ON TENSILE STRENGTH OF JACQUARD KNITTED FABRICS

BRAD Raluca, DINU Milena

Lucian Blaga University of Sibiu, Faculty of Engineering, Industrial Machinery and Equipments Department,
B-dul Victoriei 10, 550024 Sibiu, Romania, E-Mail: raluca.brad@ulbsibiu.ro

*Abstract:* *An objective approach to select the best fabric for technical and home textiles consists in mechanical properties evaluation. The goal of this study is to analyze the behavior of knitted fabrics undergoing stretch stress. In this respect, three types of 2 colors Rib structure (backstripes jacquard, twillback jacquard and double-layered 3x3 rib fabric) have been presented and tested for tensile strength and elongation on three directions. First, the elasticity and the behavior of knitted Rib fabrics were described The fabrics were knitted using 100% PAN yarns with Nm 1/15x2 on a E5 CMS 330 Stoll V-bed knitting machine, and have been tested using INSTROM 5587 Tensile Testing Machine in respect of standards conditions. After a relaxation period, 15 specimens were prepared, being disposed at 0°, 45° and 90° angles to the wale direction on the flat knitted panel. The tensile strength and the elongation values were recorded and mean values were computed. After strength and tensile elongation testing for 3 types of rib based knitted fabrics, one can see that the double layer knit presents the best mechanical behavior, followed by birds-eyebacking 2 colors Jacquard and then back striped Jacquard. For tensile stress in bias direction, the twillbacking Jacquard has a good breakage resistance value due to the higher number of rib sinker loops in structure that are positioned on the same direction with the tensile force. The twillbacking Jacquard structure could be considered as an alternative for the base material for decorative and home textile products.*

## 1. INTRODUCTION

Weft knitted structures, compared with other materials, present a purely elastic behavior, especially on transversal direction. Their elasticity, driven mainly by yarns bending, loops interlocking and the tendency to fill a minimal energy position, influences the possibilities of use. In addition, a low tightness due to inter and intraloops spaces and low mechanical properties due to fibers destruction during the knitting process, compared to other types of textile, limit the industrial knit applications [1], [2]. Nevertheless, they represent an important class of raw materials for home textiles, upholsteries and mattress sides.

The most frequent stress undergone by fabrics during wearing is traction. Tensile strain always results in a change of shape in the tensile direction. Traction forces acting on fabrics during wearing are often smaller than breaking forces. Nevertheless, they may cause irreversible damage. This kind of deformations may appear even after a first use but they are more visible after repeated stress. The size of deformations depends on stress type and duration, but it is also influenced by the raw materials used and the parameters of the knitting operation. The breaking resistance is determined by the value of the breaking force or through specific indices, through the specific resistance, the tenacity or breaking length. The traction force also depends on the fabric structure. The breaking force represents the value of the breaking force which causes the breaking of the fabric sample when applied as axial strain.

Tensile properties of weft knitted fabrics are influenced by factors such yarns interactions, structure and knitting parameters. The longer inlaid yarns on different directions increased the fabric extensibility, the thickness of the fabric and the cover factor, but reduced tensile recover [3]. The use of structures which contain floating and normal loops has benefits on tensile behavior and fabrics stiffness (elastic modulus) [4], [5]. Other researchers have tried to improve the knitted fabrics suitability for composite materials by the pre-stretch of knitted perform uni-axially and/or bi-axially before consolidation [6].

The single jersey fabric's mechanical behavior was studied in many papers and in different conditions. In this respect, [7] presents studies on the tensile properties of plain weft knitted DuPont Kevlar fiber fabric reinforced epoxy matrix composites to investigate the anisotropy of knitted fabric composites at different angles. In [8], a "cross-over model" has been proposed for expressing the cross-over of curved yarns of knitted fabric and tensile strength properties have been predicted by estimating the fracture strength of yarns bridging the fracture plane. Z. Jinyun et al. [9] describe a method for testing the elastic knitted fabric Poisson ratio and modulus was proposed based on orthotropic theory and strip biaxial tensile test. The paper [10] is focused on the bursting strength of various derivatives of single jersey knit fabric in both grey and finished state. Higher presence of tuck and miss loops in wales direction affect the bursting strength.

The mechanical properties of Rib structures are also predicted or tested in many papers. In [11], the mechanical performance of the composites which are reinforced with glass knitted fabrics composed of tuck stitches have been investigated. Tensile, compression, impact and compression-after-impact tests were performed and the results of the composites reinforced with full cardigan derivative knitted fabrics and 1×1 rib knitted fabrics of glass fibers have been compared.

The paper in [12] shows that the weft knitted fabrics with Rib structure have a superior tenacity. Also, double jersey fabrics of Full Cardigan and Full Milano represent better mechanical properties in comparison with single jersey fabrics. In [13], the Milano rib knit fabric structure has been approximated by several simpler plain stitches and its stiffness and strength of each unidirectional composite is predicted.

## 2. RIB JACQUARDS

Traditionally, quilted double-layer structure filled with unidirectional weft yarns is the base material for mattress covers. These structures are thick, with good tensile strength and dimensional stability, due to reduced elongations by the float stitch of blister yarns and weft threads. Between the double-knitted fabrics, one can emphasize the double jersey and interlock structures. Due to the spatial position of the rib sinker loop, double jersey has high elasticity on the course direction. The ratio of the transverse and longitudinal directions elongations for 1x1 Rib is approximately equal to 4.

Many weft knitted jacquards in two colors are based on 1x1 Rib fabrics. A part of front bed needles are selected to knit with first color, while those remaining are selected to knit color 2. One design row is made with 2 feeders. If all backloops are knitted with every feeder, it will result horizontally color stripes on the back of the fabric. The length of the yarn floats lies on one space needle, which will involve a small extension on course direction. The twillbacking double jersey fabric is a knitted structures obtained with one to one back needles selection. It is a more stable and balanced structure than the striped backing one. The Rib Jacquards have lower breaking deformation and tensile because of uneven traction efforts placement.

A double layer knitted fabric can be obtained on twelve needles with purl set-out. In this case, two 3x3 Rib structures are intermeshed. The knit presents a greater elasticity and smaller width than double jersey fabric. Tensile strength on course direction is two times higher than Rib fabrics one, the width and length elongation have lower values than double jersey, but not too much. The knit will not unrove from the end knitted first; it is thicker and heavier and must be made of more fines, resistant and expensive yarns. Productivity is half the rib fabric case.

## 3. MATERIALS AND METHOD

The paper presents tensile strength and strain tests on three types of knitted fabrics: horizontally back color stripes Jacquard (K1), twillbacking Jacquard (K2) and double-layered 3x3 Rib knit fabric (K3), which respected the same pattern design on the face of fabrics. The sets were knitted on a Stoll CMS 330 V-bed knitting machine, with 5 needles per inch. The yarns used in this experiment were 100% acrylic, with Nm 1/15 size. Two parallel yarns were fed to yarn feeders, which were used throughout the whole knitting process.

The knitting speed was controlled electronically at 0.9 m/min. on the same number of needles. The take-down tension (the mean value WM = 7.0) and the yarn input tension were kept constant for each type of fabric by programming the knitting machine's software. The knitted fabrics were relaxed in dry conditions for 48 hours until the structure parameters were measured, with ±(0.05 – 0.07) errors, according to British Standard BS 5441:1998 (Table 1).

*Table 1:* Dimensional values for knitted fabrics

| Fabric type | Thickness [mm.] | CPC | | WPC |
|---|---|---|---|---|
| | | front | back | |
| K1 | 2.3 | 2.14 | 4.11 | 2.14 |
| K2 | 3.6 | 3.40 | 3.30 | 2.50 |
| K3 | 4.5 | 3.61 | 3.67 | 2.64 |

For each type of fabric, tensile tests were conducted at three off-axial angles: 0°, 45° and 90°, regarding the wale direction. At least five specimens for each group with 50 mm. widths were prepared and tested on the INSTROM 5587 Tensile Testing Machine, according to ISO 1421:1998, the Strip Test Method. The distance between the clamping grips was 100 mm. and the crosshead speed was $v_1 = 15$ mm/min.

## 4. RESULTS

Figure 1 presents a view of the INSTROM machine with one sample being tested for the tensile strength. Also, three stressed specimens for each direction are shown in the second figure. It could be observed the fact that each orientation sample performed differently, due to the angles between stress and the breaking elements.

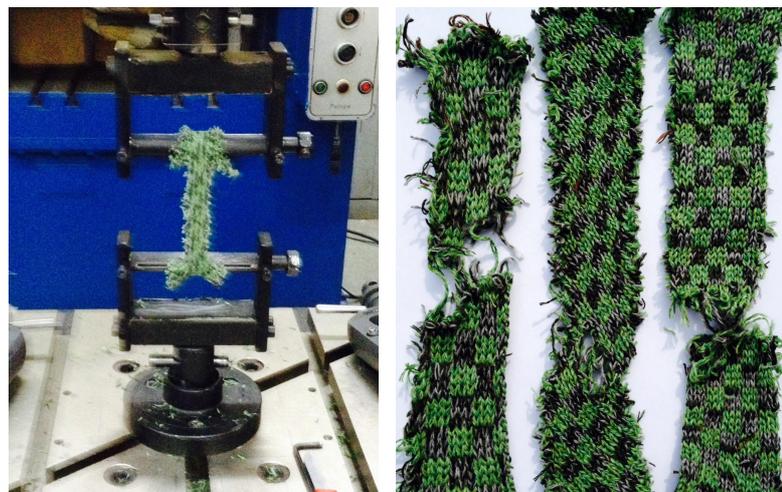

*Fig. 1*: View of the samples on the testing machine and after the stress test

*Table 2:* Mean values for stress and strain of knitted fabrics

| Fabric type | Average Stress [MPa] | Average Strain [%] |
|---|---|---|
| bias | | |
| K1 | 2.777 ± 0.197 | 38.577 ± 5.718 |
| K2 | 3.529 ± 0.284 | 46.917 ± 4.202 |
| K3 | 3.915 ± 0.232 | 57.844 ± 6.975 |
| horizontal | | |
| K1 | 1.323 ± 0.158 | 65.024 ± 9.264 |
| K2 | 2.599 ± 0.175 | 64.077 ± 6.262 |
| K3 | 3.488 ± 0.344 | 80.583 ± 6.146 |
| vertical | | |
| K1 | 4.429 ± 0.481 | 59.685 ± 5.044 |
| K2 | 4.701 ± 0.294 | 54.453 ± 3.631 |
| K3 | 5.086 ± 0.294 | 66.445 ± 2.354 |

The tensile strength values for each sample and angle were recorded. The average values were determined for all 3 knitted types, each with 5 samples, in table 2. Figure 2 shows an example for the case of K3 type and walewise direction.

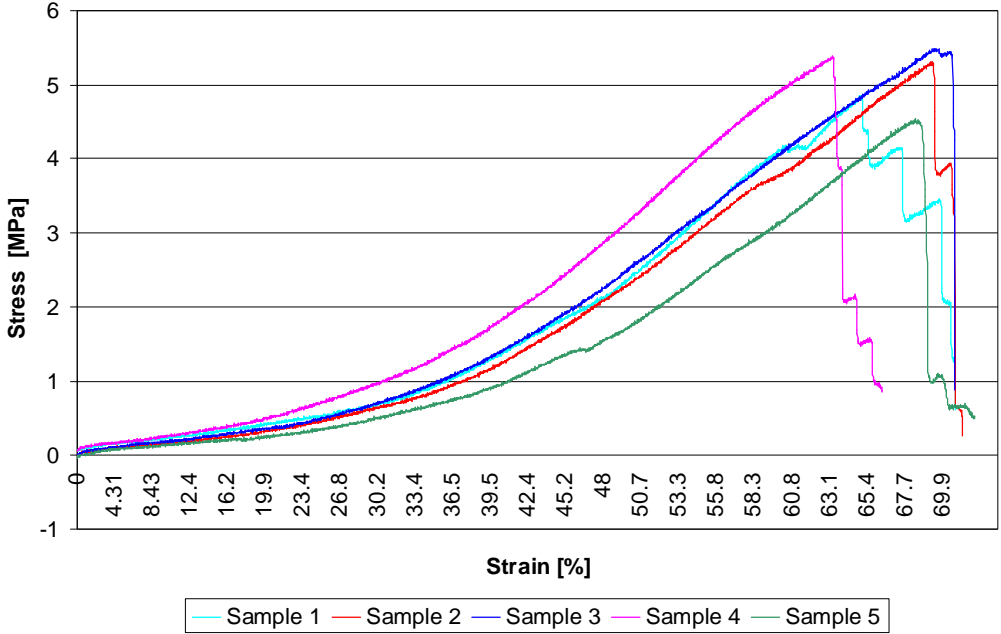

*Fig. 2: Stress – strain curve for double-layered fabric in walewise direction*

In figures 3 and 4, a comparative representation of stress and strains values of 3 fabric type and 3 tensile directions are presented.

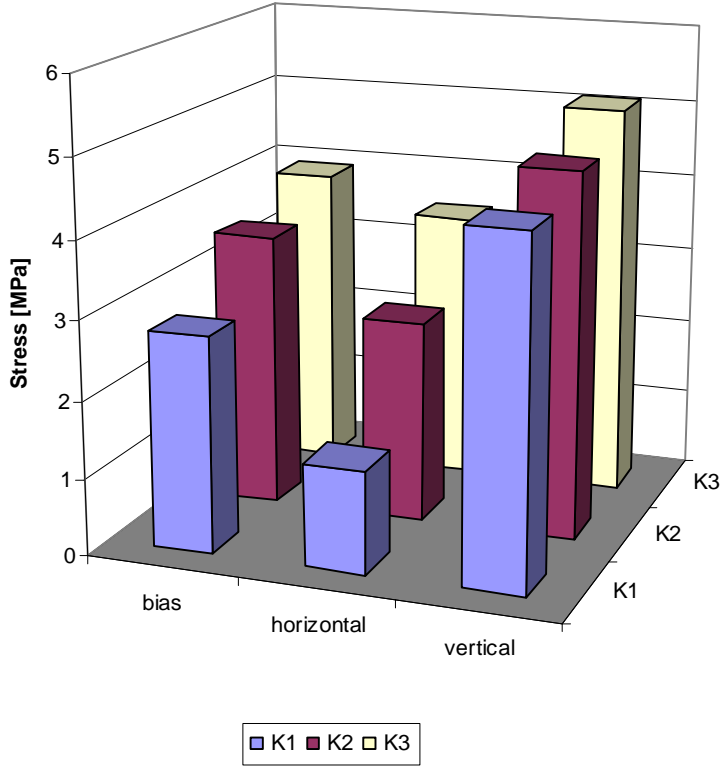

*Fig. 3: Comparative graphic of stress values of 3 rib fabric type and 3 tensile directions*

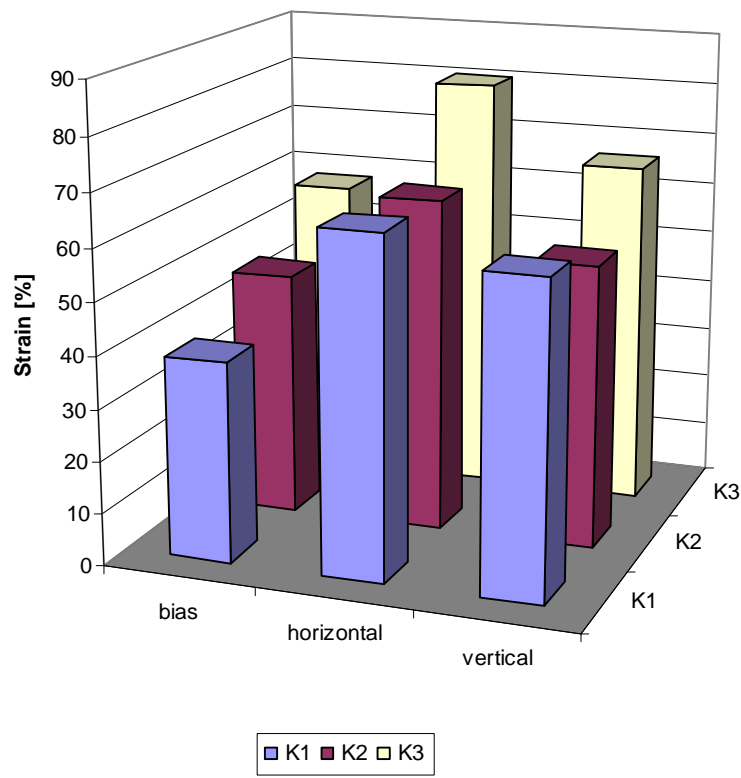

*Fig. 4: Comparative graphic of strain values*

In figure 3, one can observe that the highest values for the tensile regardless the structure corresponds to the vertical position due to a double number of side limbs in which the stresses occur, while the horizontal direction is characterized of lower values, as expected. The highest strain value was recorded for the course direction, while bias and walewise elongations are approximately the same especially for K2 and K3 samples.

## 5. CONCLUSIONS

After strength and tensile elongation testing for 3 types of 2 colors Rib based knitted fabrics, one can see that the double layer knit presents the best mechanical behavior, followed by birds-eyebacking Jacquard and then backstripes Jacquard. For tensile stress in bias direction, the twillbacking Jacquard has a good breakage resistance value due to the higher number of rib sinker loops in structure that are positioned on the same direction with the tensile force.

As a conclusion, the twillbacking Jacquard structure could be considered as an alternative for the quilted double-layer structure for the base material for mattress covers or other applications where a good resistance and elasticity are required.